\documentclass[a4paper,10pt]{scrartcl}
\usepackage{url}
\usepackage{listings}
\usepackage[a4paper,margin=3.5cm]{geometry}
\usepackage[english]{babel}
\usepackage{graphicx}
\usepackage{fancyvrb}
\usepackage{amssymb,graphicx}
\usepackage{amsmath}
\usepackage{lipsum}
\usepackage{color}
\usepackage[usenames,dvipsnames]{xcolor}
\usepackage{amssymb}
\usepackage{morefloats}
\usepackage{footnote}
\usepackage{multicol}
\usepackage{mathtools}
\usepackage{float}
\usepackage{floatflt}
\usepackage{wrapfig}
\usepackage{hyperref}
\usepackage{enumerate}
\usepackage{cleveref}
\usepackage[bf,sf]{caption}
\usepackage{heptools}
\usepackage{wrapfig,lipsum,booktabs}
\usepackage[T1]{fontenc}
\usepackage[utf8]{inputenc}
\usepackage{tikz,pgfplots}
\usetikzlibrary{math}
\usepackage{subfig}
\usepackage{scalerel}
\usetikzlibrary{math}
\usetikzlibrary{matrix,calc}
\usetikzlibrary{arrows}
\usetikzlibrary{positioning}
\usepackage{pgfplots}
\pgfplotsset{compat=1.16}
\usetikzlibrary{calc}
\usepackage[scaled=1]{helvet}
\usepackage[helvet]{sfmath}
\usepackage{amsmath,latexsym}
\usetikzlibrary{shapes.misc}
\usetikzlibrary{arrows.meta}
\usetikzlibrary{patterns,decorations.pathreplacing}
\makeatletter
\DeclareOldFontCommand{\rm}{\normalfont\rmfamily}{\mathrm}
\DeclareOldFontCommand{\sf}{\normalfont\sffamily}{\mathsf}
\DeclareOldFontCommand{\tt}{\normalfont\ttfamily}{\mathtt}
\DeclareOldFontCommand{\bf}{\normalfont\bfseries}{\mathbf}
\DeclareOldFontCommand{\it}{\normalfont\itshape}{\mathit}
\DeclareOldFontCommand{\sl}{\normalfont\slshape}{\@nomath\sl}
\DeclareOldFontCommand{\sc}{\normalfont\scshape}{\@nomath\sc}
\makeatother

\input{definitions}

\begin{document}

\title{Effective Lagrangian Morphing}
\author{Rahul Balasubramanian, Lydia Brenner, Carsten Burgard,\\ Wouter Verkerke}

\maketitle
\thispagestyle{empty}

\begin{abstract}
  In the absence of a signal of new physics being discovered at the
  \acr{LHC}, the focus on precision measurements to probe for
  deviations from the Standard Model is increasing.  One theoretically
  sound and coherent method of modelling such deviations is the use of
  effective Lagrangian density functions to introduce new terms with
  free coefficients, for example in the form of effective field
  theories. Constructing models parametric in these new coefficients is
  achieved by virtue of combining predictions from Monte Carlo
  generators for different \acr{BSM} effects. This paper builds upon
  earlier works \cite{morphing-pub,yr4} and describes a state-of-the-art
  approach to build multidimensional parametric models for new physics
  using effective Lagrangians. Documentation and tutorials
  for an associated toolkit that have been contributed to the
  \texttt{ROOT} data analysis software framework is included.
\end{abstract}

\section{Introduction}

In the absence of a discovery of new physics beyond the Standard Model
(\acr{SM}) at the LHC, the focus of many analyses has shifted towards
interpretations of precision measurements.

Standard Model Effective Field Theory (\acr{SMEFT}) provides a
theoretically elegant language to encode the new physics induced by a
wide class of beyond-the-\acr{SM} (\acr{BSM}) models that reduce to the
\acr{SM} at low energies, and is systematically improvable with
higher-order perturbative calculations. Within the mathematical
language of the \acr{SMEFT}, the effects of \acr{BSM} dynamics at high
energies $\Lambda\gg v$, well above the electroweak scale $v =
246$\,GeV, can be parametrised at low energies, $E \ll \Lambda$, in
terms of higher-dimensional operators built up from the Standard Model
fields and respecting its symmetries such as gauge invariance
\begin{equation}
  \mathcal{L}_\text{SMEFT} = \mathcal{L}_\text{SM} + \sum\limits_i^{N_{d6}} \frac{c_i}{\Lambda^2} \mathcal{O}_i^{(6)} + \sum\limits_j^{N_{d8}} \frac{b_j}{\Lambda^4} \mathcal{O}_j^{(8)} + \dots,
\label{eq:smeft}
\end{equation}
\noindent where $\mathcal{L}_\SM$ is the \acr{SM} Lagrangian,
${\mathcal{O}_i^{(6)}}$ and ${\mathcal{O}_j^{(8)}}$ represent a
complete set of operators of mass-dimensions $d = 6$ and $d = 8$, and
$c_j,b_j$ are the corresponding Wilson coefficients. Operators with $d
= 5$ and $d = 7$ violate lepton and/or baryon number conservation and
are less relevant for \acr{LHC} physics. The effective theory
expansion in Eq.\,\eqref{eq:smeft} is robust, fully general, and can be
systematically matched to explicit ultraviolet-complete \acr{BSM}
scenarios.

Measurements of cross-sections will almost always take the form of fitting a model
of the physical observations to the data. However, generally, no function
describing the physical observations from first principles can be
analytically constructed. Thus, it is necessary to construct
template distributions by means of Monte Carlo simulations of the physical 
interactions at the collision point, as well as the simulations of detector
response to the particles originating from these collisions. The
interpolation procedure using these templates to construct a function of 
the parameters that describes the data are not always based on the expected underlying physics principles but on empirical algorithms. For
nuisance parameters for example, empirical
methods of interpolating between the coordinates in parameter space at
which these templates have been generated are common. For the
parameters of interest however, more accurate methods should be
applied. In the specific case of \acr{EFT} parametrizations (or, more
generally, any Lagrangian model where the parameters take the form of
coupling strength modifiers) it is possible to construct an
analytical interpolation based on the structure of
perturbation theory and the Monte Carlo templates provided.

In recent years, the need for precise, efficient and numerically
stable implementations of such interpolations has gained significant
traction \cite{morphing-pub,yr4}. This paper sets out to present such
an implementation, capable of constructing interpolation functions to
the highest accuracy allowed by the Monte Carlo templates in terms of \acr{QCD} and \acr{EW} corrections, but also in terms of the number of \acr{EFT} injections and order of operators,
sufficiently fast to perform high-dimensional parameter fits to
extensive datasets, and versatile enough to be integrated into any
existing software framework based on \texttt{RooFit} \cite{theRooFitpackage},
the de-facto standard statistical modeling language in High Energy
Physics.

\section{Derivation}
\label{sec:derivation}
Any theory described by a Lagrangian density can be extended by
adding additional terms. In this derivation, the existing theory is
considered to be the Standard Model of Particle Physics and thus
labeled ``\acr{SM}'', but any other Lagrangian density consistent with the
observations can be used in its place. The new terms can be used to
describe new, so far unobserved interactions or particles. By nature
of this assumption, one typically considers these new interactions to
be very weak or these new particles to be very heavy in order to
explain why they have not been observed so far. Regardless, by
increasing the sensitivity of the measurement by analyzing larger
datasets or analyzing datasets taken at higher center-of-mass
energies, one might be able to observe these new effects. Typically,
however, the set of possible extensions of the existing theory is very
large, thus making it impractical to investigate them individually and
determine their compatibility with the \acr{SM} using a hypothesis test.

\subsection{Effective Field Theories}

Effective Field Theories comprise a popular and systematic approach to
to quantify possible deviations from SM in the data in terms of new 
physics operators. An effective field theory assumes the
existence of new interactions that are not allowed by the \acr{SM}
by introducing new operators with associated coupling strengths. 
The dynamics of any previously undiscovered heavy particle at high energy is captured by these operators at lower energy.
In the limit where the relevant energy scale of the experiment is much smaller than some large energy scale $\Lambda$, the Lagrangian can be treated perturbatively. Any particles with $m>\Lambda$
added to the theory are integrated out, giving rise to new interactions that will appear as some new operator $\Op_X$, with an
associated coupling $g_X$ in the Lagrangian density. As these new
interactions are not necessarily gauge interactions, effective field
theories typically do not comply with common symmetry
assumptions. Notably, the coupling strengths $g_X$ are typically not
dimensionless, causing the new interactions to exhibit nonphysical
behaviors in the high energy limit $E>\Lambda$. The dimension of these couplings
can be expressed in units of $\Lambda$ to obtain dimensionless
coefficients $c_X$ that parametrize the strength of this interaction,
e.\,g.~$g_X\cdot \Lambda^n\propto c_X$. As the number of possible
additions and extensions to any theory is of course infinite, it is
useful to categorize the new terms in the Lagrangian density function
by their suppression order $n$. In the Standard Model of Particle
Physics, where the unit of the Lagrangian density is $\unit{\Ell} =
\GeV^4$, the suppression order $n$ is directly related to the
dimension $d$ of the new operator $O_X$ as $n+4=d$. For the
interactions studied at the \acr{LHC} experiments, operators with odd
dimensionality are typically not considered, as they would violate
flavour number conservation, such that the first ``interesting'' terms
appear at $d=6$, and further at $d=8$.

-\subsection{From Lagrangians to cross sections}

In High Energy Physics, the most common type of quantity reported is a
cross section, a measure of the probability of a specific process
transforming an initial state $i$ into a final state $f$, typically
measured in barn. Considering an underlying theory with the Lagrangian density
\begin{align}
  \Ell = g_\SM\Op_\SM + g_\BSM\Op_\BSM, \label{eq:smeft-lagrangian}
\end{align}
where $g_\SM$ and $g_\BSM$ are the \acr{SM} and \acr{BSM} coupling
strengths, and $\Op_\SM$ and $\Op_\BSM$ the corresponding operators
describing the standard model and new physics processes. The cross
section $\sigma_{i\to f}$ of some process can be obtained from the
phase space integral of the matrix element \ME{} squared,
\begin{align}
  \sigma_{i\to f} &\propto \int  \abs{\ME_{i\to f}}^2.
\end{align}

\begin{figure}[htb]
  \hfill  
  \begin{tikzpicture}
  \coordinate[bbvertex] (v) at(0,0);
  \draw[] (-1,1) -- (v);
  \draw[] (-1,-1) -- (v);
  \draw[] (v) -- (1.4,0);    
\end{tikzpicture}
  \hfill
  \begin{tikzpicture}
  \coordinate[bbvertex] (v) at(0,0);
  \draw[] (1,1) -- (v);
  \draw[] (1,-1) -- (v);
  \draw[] (v) -- (-1.4,0);    
\end{tikzpicture}
  \hfill  
  \begin{tikzpicture}
  \coordinate[bbvertex] (v1) at(0,0);
  \coordinate[bbvertex] (v2) at(1.4,0);  
  \draw[] (-1,1) -- (v1);
  \draw[] (-1,-1) -- (v1);
  \draw[] (v1) -- (v2);    
  \draw[] (2.4,1) -- (v2);
  \draw[] (2.4,-1) -- (v2);
\end{tikzpicture}
  \hfill~  \par
  \caption{Feynman diagrams of a single-vertex particle production (left) or decay (center) as well as for a $2\to2$ $s$-channel process (right). The vertices are depicted as large circles with a fill pattern (\protect\tikz{\protect\coordinate[scale=0.5,bbvertex] (v) at(0,0);}) to indicate their nature as ``effective'' vertices.\label{fig:feynman}}
\end{figure}
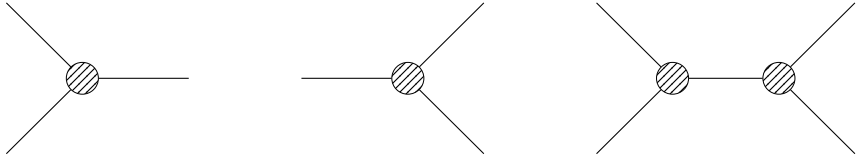

The matrix element $\ME_{i\to f}$ is in turn obtained by applying the Feynman rules to all diagrams
describing this process. This will always lead to a sum of products of individual terms from the Lagrangian density, e.\,g.
\begin{align}\label{eq:me-decomposed}
   \abs{\ME_{i\to f}}^2 &= g^2_\SM \Op^2_\SM + g_\SM g_\BSM \Op_\SM \Op_\BSM + g^2_\BSM \Op^2_\BSM.
\end{align}
Eq.\,\eqref{eq:me-decomposed} is the parametrization of the matrix element for a single-vertex
diagram, such as the production or the decay of a Higgs boson, as illustrated by the first two Feynman diagrams shown in Figure \ref{fig:feynman}.
The corresponding cross section in any kinematic phase space can be written as
\begin{align}\label{eq:xsec-decomposed}
  \sigma_{i\to f} &\propto&& g^{2}_\SM \int  \Op^2_\SM
  + g_\SM g_\BSM \int  \Op_\SM \Op_\BSM 
  + g^2_\BSM \int  \Op^2_\BSM,
\end{align}  
where the integrals no longer depend on the values of the couplings
and can be evaluated using standard Monte Carlo techniques. As the
couplings $g$ are suppressed by $\Lambda^{-2}$ for dimension-6
operators and by $\Lambda^{-4}$ for dimension-8 operators, etc., the
expression can be ordered in powers of $\Lambda$. A prediction for a fixed power in $\Lambda$ can be obtained by terminating the sequence at any given
point. In practice, due to the limited availability of Monte Carlo
generators with higher-dimensional operators, most processes can only
be modelled consistently with terms up to and
including $\Lambda^{-2}$, whereas any higher order in $\Lambda$
cannot be calculated yet fully due to the incomplete availability of
implementations of dimension-8 operators, which enter at the order of $\Lambda^{-4}$, in Monte Carlo generators.

This derivation easily generalizes to a $2 \to 2$, $s$-channel Feynman diagram, such as the one shown in right in Figure \ref{fig:feynman}. In this case, the matrix element reads
\begin{align}\label{eq:me-decomposed-2to2}
   \abs{\ME_{i\to f}}^2 &= g^4_\SM \Op^4_\SM + g^{3}_\SM g_\BSM \Op^{3}_\SM \Op_\BSM + g^{2}_\SM g^2_\BSM \Op^{2}_\SM \Op^2_\BSM + g_\SM g^3_\BSM \Op_\SM \Op^3_\BSM + g^4_\BSM \Op^4_\BSM 
\end{align}
and the cross-section relation in any kinematic phase space is given by,
\begin{align}\label{eq:xsec-decomposed-2to2}
  \sigma_{i\to f} &\propto&& g^{4}_\SM \int  \Op^4_\SM \notag\\&&&  
  + g^{3}_\SM g_\BSM \int  \Op^{3}_\SM \Op_\BSM \notag\\&&&
  + g^{2}_\SM g^2_\BSM \int  \Op^{2}_\SM \Op^2_\BSM \\&&&  
  + g_\SM g^{3}_\BSM \int  \Op_\SM \Op^{3}_\BSM \notag\\&&&  
  + g^4_\BSM \int  \Op^4_\BSM \notag,
\end{align}\noindent
In this section, only one \acr{BSM} operator $\Op_\BSM$ with an
associated coupling $g_\BSM$ was considered. However, when considering
more operators $\Op_X,\Op_Y,\dots$, this only changes the number of
terms appearing in Equation \eqref{eq:xsec-decomposed}, not the
structure of the equation. The additional bookkeeping of operators and
powers involved in these equations is provided by the implementation
of \texttt{RooLagrangianMorphFunc}.

\subsection{Effective Lagrangian Morphing}
\label{sec:elm}

Equation \eqref{eq:xsec-decomposed} or any truncation thereof
is nothing but a linear combination of predicted cross section components
$\sigma_\star$ in the form of individual phase space integrals $\int
\Op_X\Op_Y\dots$, with the coefficients taking the form of
polynomials in the couplings $g_X,g_Y,\dots$. This observation opens
up the possibility of describing the prediction $\sigma_{i\to f}$ as a
function of the vector of couplings $\vec{g}$ with the help of
individual pre-calculated phase space integrals (templates, created for some fixed \acr{EFT} scenarios $A$, $B$, \dots).  These
integrals can be pre-computed at truth level or, crucially, at the detector level if the simulation is extended to include the detector simulation. The basis of the phase space integral is mapped to the available predictions $\{A,B,\dots\}$ with the morphing matrix $M$ whose elements are composed of the coupling polynomial that scale each of the phase space integral.

The values of the coefficient $g_i$ for which these templates are
generated do not need to follow a rigid grid-like pattern  where individual
couplings are chosen at a value of either 0 or 1.
Some Monte Carlo generators
allow to separately estimate allow to generate each of the integrals that enter Eq.\,\eqref{eq:me-decomposed}. In this case, the individual templates themselves
do not correspond to physically viable scenarios but constructing the function
$\sigma(\vec{g})$ is straight-forward. The morphing matrix $M$ in this case is fully diagonal
\begin{align} M &= \begin{pmatrix}
  g^{m}_{X,A} &  & \\
  & g^{m-1}_{X,B} g^{1}_{Y,B} & \\
  &  & \ddots \\
  \end{pmatrix}
  \label{eq:morph_matrix-diag}
\end{align}

In the general case that generators can only provide physically meaningful scenarios, and not the isolated terms, the morphing matrix can still be constructed from such samples, but the matrix will acquire non-diagonal elements in that case, and it is important that sampling points in the parameter space are chosen in such a way that the matrix is invertible.
The phase space integrals can be
inferred from the provided templates by linear algebra: If the expression for
$\sigma_{i\to f}$ involves $m$ different polynomials, a set of $m$
different predictions of $\sigma_{i\to f}$ at arbitrary coordinates
$\vec{g}_A,\vec{g}_B,\dots$, can be used to compute the set of
$\sigma_\star$ by virtue of the morphing matrix $M$ constructed as
\begin{align} M &= \begin{pmatrix}
  g^{m}_{X,A} & g^{m-1}_{X,A} g^{1}_{Y,A} & \cdots & g^{1}_{X,A} g^{m-1}_{Y,A} & g^m_{Y,A}\\
  g^{m}_{X,B} & g^{m-1}_{X,B} g^{1}_{Y,B} & \cdots & g^{1}_{X,B} g^{m-1}_{Y,B} & g^m_{Y,B}\\
  \vdots & \vdots & & \vdots & \vdots
  \end{pmatrix}
  \label{eq:morph_matrix}
\end{align}\noindent

Again, as the number of expressions in Equation
\eqref{eq:xsec-decomposed} grows larger, also the morphing matrix
increases in dimensionality, as the combinatorial possibilities of
constructing unique polynomials of order $m$ from a vector of
couplings $\vec{g}$ increases, but no conceptual difficulty arises. On
the other hand, if Equation \eqref{eq:xsec-decomposed} was truncated
at a certain order in \acr{EFT}, the morphing matrix $M$ loses the
corresponding columns, and the number of samples can be reduced
accordingly.
The complete morphing function $\sigma_{i\to
  f}\left(\vec{g}\right)$, which provides a description of the cross sections for any value of the couplings in terms of the available predictions $\sigma_A,\sigma_B,\dots$, where $\sigma$ denotes the cross-section of the kinematic distribution of an observable, as
\begin{align}
  \sigma_{i\to f}\left(\vec{g}\right) &= M^{-1} \cdot \begin{pmatrix}\sigma_A\\ \sigma_B\\ \vdots\end{pmatrix}, \label{eq:morphfunc}
\end{align}
as long as the morphing matrix $M$ is invertible, which can always be
achieved as the couplings $\vec{g}$ can be chosen to not be a degenerate set
when generating the samples.

This approach is capable of extracting the cross section components
$\sigma_{\star}$ from any sufficiently powerful set of pre-computed
phase-space integrals: for explicitly generated single integrals,
corresponding to a fully diagonal morphing matrix, as well as for
integrals with mixtures of different contributions as well as any
combination. 

\subsection{Morphing models in Higgs physics}

The Higgs boson is predicted to be a narrow-width (that is,
short-lived and predominantly on-shell), CP-even particle with spin
$s=0$, with no contrary evidence discovered so far \cite{atlas-hwwcp,atlas-hyycp,atlas-httcp,cms-tthcp,cms-httcp}. The dominant
Feynman diagrams involving the Higgs boson at the \acr{LHC} can thus
be separated into two separate components: the Higgs boson
\textit{production} process $i\to H$, typically from two incoming particles, and the
Higgs \textit{decay} process $H\to f$, typically into two outgoing particles. Due to
the narrow width and the vanishing quantum numbers, no information can
be transferred between the production and the decay vertex other than
the four-momenta $\pvec{p}_H$ of the Higgs boson. The cross section prediction of Higgs boson
production and decay thus in very good approximation
factorizes into the prediction of the cross section for Higgs boson
production $\sigma_{i\to H}$, as a function of its four-momenta, and
the branching ratio $\BR_{H\to f}$ of the Higgs boson decay, which can
be evaluated in its rest frame as the ratio of the partial decay width
$\Gamma_{H\to f}$ and total width of the Higgs boson $\Gamma_H$:
\begin{align}\label{eq:factorization}
  \sigma_{i\to H\to f}(\pvec{p}_H) &\approx \sigma_{i\to H}(\pvec{p}_H) \cdot \BR_{H\to f} = \sigma_{i\to H}(\pvec{p}_H) \cdot \frac{\Gamma_{H\to f}}{\Gamma_H} 
\end{align}
The decay widths $\Gamma_{H\to f}$ and $\Gamma_H$ are, similar to
cross-sections, directly related to the matrix element and can be
modeled using Effective Lagrangian Morphing to include effects of 
hypothetical new physics. In summary, both the complete process 
$\sigma_{i\to H\to f}$ as well as the individual components 
$\sigma_{i\to H}$, $\Gamma_{H\to f}$ and $\Gamma_H$ can be expressed 
as polynomials of the couplings $\vec{g}$ and some pre-calculated 
predictions using Effective Lagrangian Morphing.

If the statistical model used to infer the data performs an unfolding
to truth particle level, for example through the procedure of
simplified template cross-sections \cite{stxs}, the parametrization of
the cross section can happen on truth level, while the data and the
background predictions enter the likelihood at detector level. In such
a case, the factorized cross section prediction at detector level can
be decomposed into components including the elements of the confusion
matrix $\epsilon_{ij}$ and the truth-level signal predictions derived
by the morphing procedure. Concretely, for a number of events $N_i$ in
any bin $i$ of the measurement,
\begin{align}
  N^\reco_i &= N_i^{\bkg,\reco} + N_i^{\sig,\reco}\\
  &= N_i^{\bkg,\reco} + \sum_j \epsilon_{ij} N^{\sig,\truth}_j\\
  \label{eq:stxs}  
  &= N_i^{\bkg,\reco} + \sum_j \epsilon_{ij} \frac{1}{\prime} \sigma_{j} \cdot \BR.
\end{align}
Here, $\mathcal{L}^\prime$ is the Luminosity for which the
predictions have been generated, $\sigma_j$ is the Higgs boson
production cross section for truth bin $j$, and $\epsilon_{ij}$ is the
confusion matrix expressing the fraction events in truth bin $j$ that
end up being measured in bin $i$ at detector level.

\subsection{Linearization}
\label{sec:linear}
The truth-level signal prediction $\sigma \cdot \BR$ is typically
considered in the form of a \acr{LO} or \acr{NLO} multiplicative correction factor
for the \acr{SM} prediction calculated at the highest available order
(e.\,g.~up to \acr{N$^{3}$LO}), such that

\begin{align}
  \sigma^{i\to H} &= \sigma^{\prime i\to H}_\SM \cdot \left ( 1 + \sum_k \frac{c^{(6)}_k}{\Lambda^2}\frac{\sigma^{i\to H}_{k}}{\sigma^{i\to H}_\SM} + \sum_{kl} \frac{c^{(6)}_k}{\Lambda^2} \frac{c^{(6)}_l}{\Lambda^2} \frac{\sigma^{i\to H}_{kl}}{\sigma^{i\to H}_\SM} + \dots \right) \label{eq:prod}\\
  \Gamma^{H\to f} &= \Gamma^{\prime H\to f}_\SM \cdot \left ( 1 + \sum_k \frac{c^{(6)}_k}{\Lambda^2}\frac{\Gamma^{H\to f}_{k}}{\Gamma^{H\to f}_\SM} + \sum_{kl} \frac{c^{(6)}_k}{\Lambda^2} \frac{c^{(6)}_l}{\Lambda^2} \frac{\Gamma^{H\to f}_{kl}}{\Gamma^{H\to f}_\SM} + \dots \right) \label{eq:partial_width}\\
  \Gamma &= \Gamma^{\prime H}_\SM \cdot \left ( 1 + \sum_k \frac{c^{(6)}_k}{\Lambda^2}\frac{\Gamma^{H}_{k}}{\Gamma^{H}_\SM} + \sum_{kl} \frac{c^{(6)}_k}{\Lambda^2} \frac{c^{(6)}_l}{\Lambda^2} \frac{\Gamma^{H}_{kl}}{\Gamma^{H}_\SM} + \dots \right) \label{eq:total_width}   
\end{align}

Here, the cross sections and widths used as a \acr{SM} reference calculation
($\sigma^{\prime i\to H}_\SM$, $\Gamma^{\prime H\to f}_\SM$,
$\Gamma^{\prime H}_\SM$) do not necessarily need to match the
predictions used for the \acr{SMEFT} corrections ($\sigma^{i\to H}_\SM$,
$\Gamma^{H\to f}_\SM$, $\Gamma^{H}_\SM$). They could employ a
different generator, or use a higher precision calculation (such as
NNLO instead of LO), the latter being common practice to achieve the
highest precision parametrizations (see, e.\,g., Ref.~\cite{atlas-hcomb-eft}).

Inserting Eq.\,(\ref{eq:prod}-\ref{eq:total_width}) in Eq.\,\eqref{eq:factorization} leads to the rather unwieldy expression
\begin{align}
  \sigma \cdot \BR =
  \sigma^{\prime i\to H}_\SM \cdot
  \frac{\Gamma^{\prime H\to f}_\SM}{\Gamma^{\prime H}_\SM} &\cdot 
  \left ( 1 + \sum_k \frac{c^{(6)}_k}{\Lambda^2}\frac{\sigma^{i\to H}_{k}}{\sigma^{i\to H}_\SM} + \sum_{kl} \frac{c^{(6)}_k}{\Lambda^2} \frac{c^{(6)}_l}{\Lambda^2} \frac{\sigma^{i\to H}_{kl}}{\sigma^{i\to H}_\SM} + \dots \right) \label{eq:ratiopoly} \\
  &\cdot\frac{ 1 + \sum_k \frac{c^{(6)}_k}{\Lambda^2}\frac{\Gamma^{H\to f}_{k}}{\Gamma^{H\to f}_\SM} + \sum_{kl} \frac{c^{(6)}_k}{\Lambda^2} \frac{c^{(6)}_l}{\Lambda^2} \frac{\Gamma^{H\to f}_{kl}}{\Gamma^{H\to f}_\SM} + \dots }
  { 1 + \sum_k \frac{c^{(6)}_k}{\Lambda^2}\frac{\Gamma^{H}_{k}}{\Gamma^{H}_\SM} + \sum_{kl} \frac{c^{(6)}_k}{\Lambda^2} \frac{c^{(6)}_l}{\Lambda^2} \frac{\Gamma^{H}_{kl}}{\Gamma^{H}_\SM} + \dots }\nonumber
\end{align}\noindent
and introduces some complication in the power counting, as a consistent
interpretation would require all terms of up to a given order to be included. For the interpretability of the result it is often desirable to truncate the expansion at some given order, in order to exclude contributions from higher order terms for which calculations are not fully available. Simplification of
Eq.\,\eqref{eq:ratiopoly} can be achieved by either truncating the expression 
at a fixed power of $\Lambda$ or by truncating the expression at fixed number
of \acr{EFT} operator insertions. The current implementation of \texttt{RooLagrangianMorphFunc} truncates Eq.\,\eqref{eq:ratiopoly} based on the former corresponding to a fixed order in powers of $\Lambda$. 

Truncating the Taylor expansion of Eq.\,\eqref{eq:ratiopoly} around $c_i=0$ to $\Lambda^{-2}$ reduces this to an expression of the form
\begin{align}
  \label{eq:linearized-noacc}
  \sigma \cdot \BR &\approx
  \sigma^{\prime i\to H}_\SM \cdot
  \frac{\Gamma^{\prime H\to f}_\SM}{\Gamma^{\prime H}_\SM} \cdot
  \left ( 1 + \sum_k \frac{c^{(6)}_k}{\Lambda^2} \left(\frac{\sigma^{i\to H}_{k}}{\sigma^{i\to H}_{\SM}}  + \frac{\Gamma^{H\to f}_{k}}{\Gamma^{H\to f}_\SM} -\frac{\Gamma^{H}_{k}}{\Gamma^{H}_\SM} \right) \right).
\end{align}\noindent
With only dimension-6 operators being completely available in the present state of \acr{SMEFT} simulation, this linear form provides a consistent interpretation with all available terms up to order $\Lambda^{-2}$ included.

To also include terms with dimension-6 operators squared, suppressed at power $\Lambda^{-4}$, a second order Taylor expansion can be used. While missing the terms linear in dimension-8 operators, which are also suppressed at power $\Lambda^{-4}$, these higher-order expansion are commonly used to test the sensitivity of analyses to the effect of such higher order operators. 
The second order contribution with only dimension-6 modifications is given by,
\begin{align}
\sum_{k,l}^{k\neq l} \Bigg(&\frac{c^{(6)}_k}{\Lambda^2} \Bigg)^2 \Bigg( \frac{\Gamma^{H\to f}_{kk}}{\Gamma^{H\to f}_\SM} -  \frac{\Gamma^{H}_{kk}}{\Gamma^{H}_\SM} + \frac{\sigma^{i\to H}_{kk}}{\sigma^{i\to H}_\SM} - \frac{\Gamma^{H\to f}_{k}}{\Gamma^{H\to f}_\SM} \frac{\Gamma^{H}_{k}}{\Gamma^{H}_\SM} +\left(\frac{\Gamma^{H}_{k}}{\Gamma^{H}_\SM}\right)^2 + \frac{\Gamma^{H\to f}_{k}}{\Gamma^{H\to f}_\SM} \frac{\sigma^{i\to H}_{k}}{\sigma^{i\to H}_\SM} - \frac{\Gamma^{H}_{k}}{\Gamma^{H}_\SM} \frac{\sigma^{i\to H}_{k}}{\sigma^{i\to H}_\SM} \Bigg) \notag\\
+ &\frac{c^{(6)}_k}{\Lambda^2}\frac{c^{(6)}_l}{\Lambda^2} \Bigg( \frac{\sigma^{i\to H}_{kl}}{\sigma^{i\to H}_\SM}  + \frac{\sigma^{i\to H}_{kl}}{\sigma^{i\to H}_\SM} - \frac{\Gamma^{H}_{kl}}{\Gamma^{H}_\SM} - \frac{\Gamma^{H}_{k}}{\Gamma^{H}_\SM} \frac{\Gamma^{H\to f}_{l}}{\Gamma^{H\to f}_\SM} + \frac{\sigma^{i\to H}_{k}}{\sigma^{i\to H}_\SM}\frac{\Gamma^{H\to f}_{l}}{\Gamma^{H\to f}_\SM} - \frac{\Gamma^{H\to f}_{k}}{\Gamma^{H\to f}_\SM} \frac{\Gamma^{H}_{l}}{\Gamma^{H}_\SM} \notag \\
& \ \ \ \ \ \ \ \ \ \ \ + 2 \frac{\Gamma^{H}_{k}}{\Gamma^{H}_\SM} \frac{\Gamma^{H}_{l}}{\Gamma^{H}_\SM} - \frac{\sigma^{i\to H}_k}{\sigma^{i\to H}_\SM} \frac{\Gamma^{H}_{l}}{\Gamma^{H}_\SM}
+ \frac{\Gamma^{H\to f}_{k}}{\Gamma^{H\to f}_\SM}\frac{\sigma^{i\to H}_{l}}{\sigma^{i\to H}_\SM} - \frac{\Gamma^{H}_{k}}{\Gamma^{H}_\SM} \frac{\sigma^{i\to H}_{l}}{\sigma^{i\to H}_\SM} \Bigg). 
\end{align}
The second-order expansions is also supported by the \rlmf software.

\subsection{Acceptance}
\label{sec:acceptance}

When plugging Eq.\,\eqref{eq:linearized-noacc} back into
Eq.\,\eqref{eq:stxs}, one potential shortcoming becomes apparent: The elements
$\epsilon_{ij}$ of the confusion matrix might also depend on the Wilson
coefficients $c$. Including such effects requires an additional explicit parameterization of the expression $\epsilon_{ij}$ used in Eq.\,\eqref{eq:stxs}. For any (fiducial) cross-section the product of acceptance and efficiency will always depend on the signal properties, which may be modified by the onset of \acr{BSM} operators and thus on the assumed strengths of the \acr{BSM} operator couplings. The signal extraction strategy (such as a likelihood fit) will make assumptions on signal kinematics which can again be modified due to onset of \acr{BSM} operators.
While these effects are often small enough to be ignored, they should
be quantified and accounted for at least in the case of precision
measurements. In the example of the \acr{STXS} Higgs analysis regions,
the binning of these regions has been chosen with the intent of
minimizing these issues.

It should be noted that when detector-level templates are used, the
acceptance and efficiency effects are by definition taken into account
already, as the \acr{BSM} samples have undergone detector
simulation. Only when the factorization approach from
Sec.~\ref{sec:linear} is used, the acceptance and efficiency
corrections need to be parameterized explicitly. As the total width
does not depend on the analysis selection, only changes in partial
width $\Gamma^{H\to f}_{k},\Gamma^{H\to f}_{kl}$ need to be considered. In the example of 
Ref.~\cite{atlas-hcomb-eft}, only the partial width of $H\to 4\ell$ was
handled explicitly, as for two-body-decays of the Higgs boson, there are no
kinematic degrees of freedom in the decay. These acceptance corrections can 
take the form of arbitrary functions that can be inferred from simulation 
data, and including them in the prediction using the \rlmf software is straight-forward.

\section{RooFit implementation}

The Effective Lagrangrian Morphing framework is implemented within the
\rf library \cite{theRooFitpackage}. \rf is an object-orient language 
to describe probability models and is widely used in high-energy physics.
It is available within the \texttt{ROOT} software which is an open-source data 
analysis framework used in high energy physics and other domains. The
\rf language allows for users to construct probability models of 
arbitrary complexity by expressing the relation between observables
and probability functions as an expression tree of \rf objects by 
designing all the relevant mathematical concepts as \rf classes. 
The \rlmf software that is discussed in the following section is included
from \texttt{ROOT} release v6.26 onwards. 
\begin{figure}[h!]
\centering
\subfloat{\includegraphics[width=1.0\linewidth]{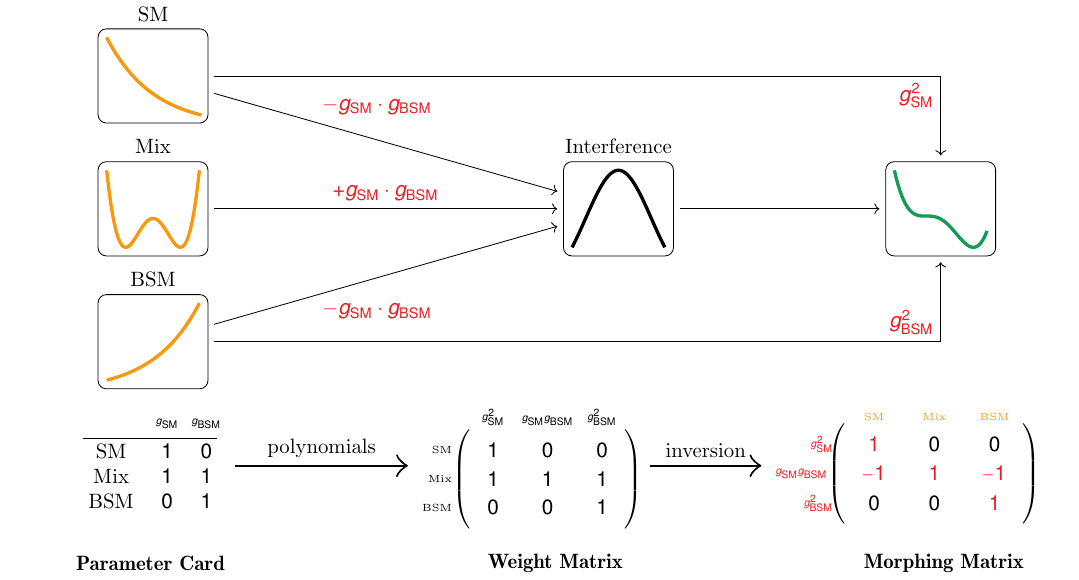}}\\
\subfloat{\includegraphics[width=1.0\linewidth]{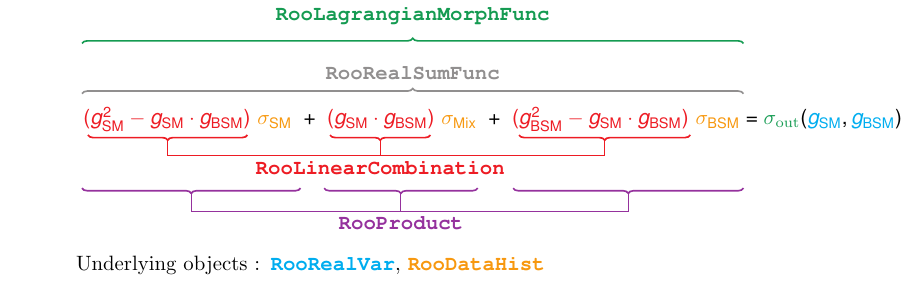}}
\caption{Schema showing the design of \texttt{\rlmf} class in \rf for a simple case involving and effective Lagrangian with two parameters $g_{\SM}$ and $g_{\BSM}$. The cross section $\sigma_{\text{out}}$ is a distribution of an observable and defined for any ($g_{\SM},g_{\BSM}$) in terms of the input templates of the observable denoted by $\{\sigma_{\SM},\sigma_{\text{Mix}},\sigma_{\BSM}\}$. In the above, the observable distributions correspond to $(g_{\SM},g_{\BSM})=\{(1,0),(1,1),(0,1)\}$ respectively.}
\label{fig:morph_design}
\end{figure}

The \rlmf class in \rf implements the Effective Lagrangian Morphing
method, as derived in Section \ref{sec:derivation}. The morphing
distribution can be constructed for an arbitrary number of parameters
$\vec{g}$ as long as the required number of non-degenerate samples are
provided as an input to the morphing function. The morphed
distribution provides a continuous description of the observable
distribution in the parameter space, as spelled out in
Eq.\,\eqref{eq:morphfunc}. The \texttt{RooLagrangianMorphFunc}
implements the morphing as a sum of functions where each of the
function is given by a \texttt{RooLinearCombination} object. The
\texttt{RooLinearCombination} class implements the underlying
summation of weight involved for each template as a Kahan sum to
reduce loss of numeric precision that may occur in the repeated
addition of a large number of summation terms.

\begin{figure}[h!]
\centering
\includegraphics[width=0.3\textwidth]{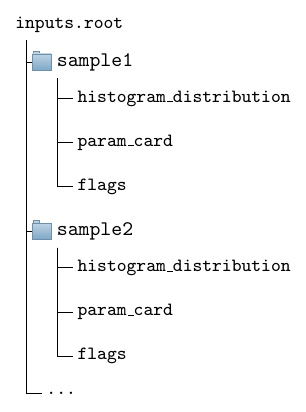}
\vspace{-12pt}
\caption{Expected structure of input \texttt{ROOT} file provided to the morphing function}
\label{fig:folder}
\end{figure}

To streamline the creation of a morphing function, the relevant 
configuration required are provided by the user through a \rlmfconfig
object  to configure the observable name, necessary parameters, 
coupling structures, and the input templates required to define 
a morphing function. Once created, the morphing function provides
a description of the observable distribution for any coordinate 
in the parameter space.

The user is required to provide non-degenerate distributions of 
the observable at a set of coordinates (corresponding to $\sigma_A$,
$\sigma_B$, \dots in Eq.\,\eqref{eq:morphfunc}) and the corresponding
parameter card which is used by the class to construct the morphed 
distribution in terms of the parameters $\vec{g}$. The input data are
expected to be structured in the manner shown in Fig.~\ref{fig:folder},
where \texttt{sample1}, \texttt{sample2}, and so on correspond to the
different input folders (\texttt{TFolder}) and the \texttt{histogram\_distribution} 
represents the template distribution that is picked up the morphing function (\texttt{TH1}).
The \texttt{param\_card}s (\texttt{TH1} with labeled axis) contain the truth parameter values
for the corresponding sample. The axis label correspond to the parameter names and the entry for the corresponding parameter is the truth value. The \texttt{flags} (\texttt{TH1} with bin labels) contain the information of powers of $\Lambda^{-2}$ that is included in the simulated sample. The \texttt{flags} axis is given by \{\texttt{nNP0}, \texttt{nNP1}, \texttt{nNP2}, ..\} which corresponds to parameterss to track the contributions of $\Lambda^{-2}$. This is used to represent if the simulated sample is solely comprised of the \acr{SM} contribution \{\texttt{nNP0}=1, \texttt{nNP1}=0, \texttt{nNP2}=0\}, \acr{SM}-\acr{BSM} interference \{\texttt{nNP0}=0, \texttt{nNP1}=1, \texttt{nNP2}=0\}, \acr{BSM}-\acr{BSM} interference \{\texttt{nNP0}=0, \texttt{nNP1}=0, \texttt{nNP2}=1\}, or a mix of them \{\texttt{nNP0}=1, \texttt{nNP1}=1, \texttt{nNP2}=1\}.

\subsection{Usage}

\begin{figure}[h!]
    \centering
    \includegraphics[width=1.05\textwidth]{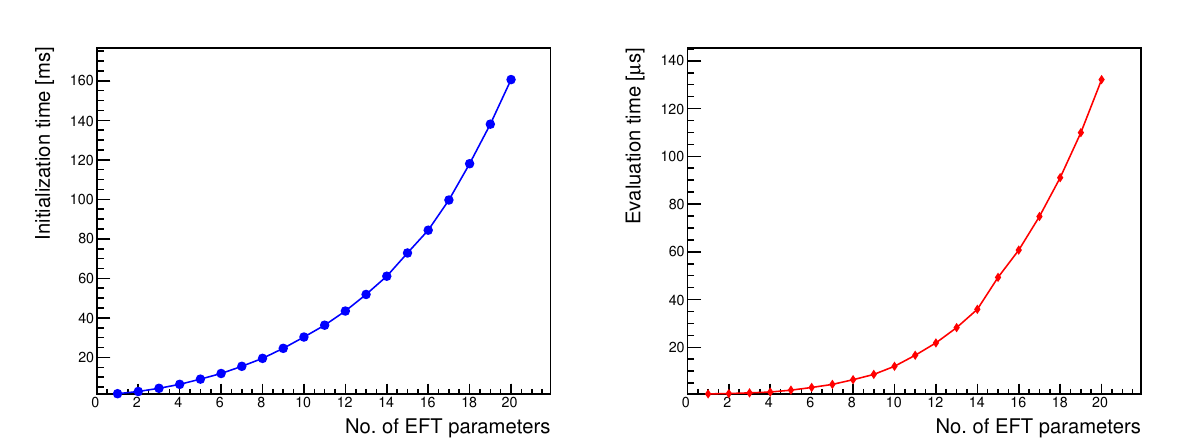}
    \caption{The dependence of the initialization and evaluation \acr{CPU} time as a function of number of parameters of the morphing function. The estimate for the initialization time corresponds to the median value of the initialization called executed 5000 times. The evaluation time corresponds to the median of 5000 calls to the morphing function evaluation at randomly chosen parameter values.}
    \label{fig:morph_time}
\end{figure}

The construction of a morphing function is efficient with the most 
expensive steps in typical cases being the matrix inversion of the 
morphing matrix as represented in Eq.\,\eqref{eq:morph_matrix} to obtain
the weight matrix and the I/O of the \texttt{ROOT} file containing the input templates. This inversion needs to be done only once at the initialization phase.
Here both the linear and quadratic \acr{BSM} operator terms are
considered and hence the number of terms in the polynomial scales as 
given in Eq.\,\eqref{eqn:nsamples}. The dependence of initializing a 
morphing function and the evaluation time of the morphing function 
with the number of \acr{EFT} parameters is shown in Fig.~\ref{fig:morph_time}.

The number of input template distributions required depends on the 
number of terms in the polynomial relation. For $n$ parameters with
single insertion of \acr{EFT} operators in the Feynman diagram the total
number of samples $N$, that is required to construct the morphing
distribution is given by,

\begin{equation}
 N =
    \begin{cases}
      n+1 & \text{, for linear terms in the \acr{BSM} operators}\\
      \ffrac{n^2+3n+2}{2} & \text{, for linear \& quadratic \acr{BSM} operator terms}\\
    \end{cases}
\label{eqn:nsamples}
\end{equation}\noindent
To construct the morphing function C++ object, all input templates
and their sampling coordinates in the parameter space must be specified 
in one transaction to the constructor. For ease of use, the \texttt{RooLagrangianMorphFunc}
object can be configured with a \texttt{RooLagrangianMorphFunc::Config} which 
can be initialized step-by-step by the user adding one at a time the relevant couplings,
input sample names, and the path to the \texttt{ROOT} file containing the
input templates to the morphing function as shown in the example \ref{lst:1d}. Alternatively, the morphing function can be constructed through through the \texttt{RooWorkspace}
factory language, as shown in the example \ref{lst:1dfac} where a named-argument syntax compared to that of other \texttt{RooFit} operator classes can be used to structure the input information. The snippets for the example usage of the morphing function with a single and multi-parameter use case are taken from the tutorials available within \texttt{ROOT}.

\begin{figure}[h!]
    \centering
    \includegraphics[width=1.0\textwidth]{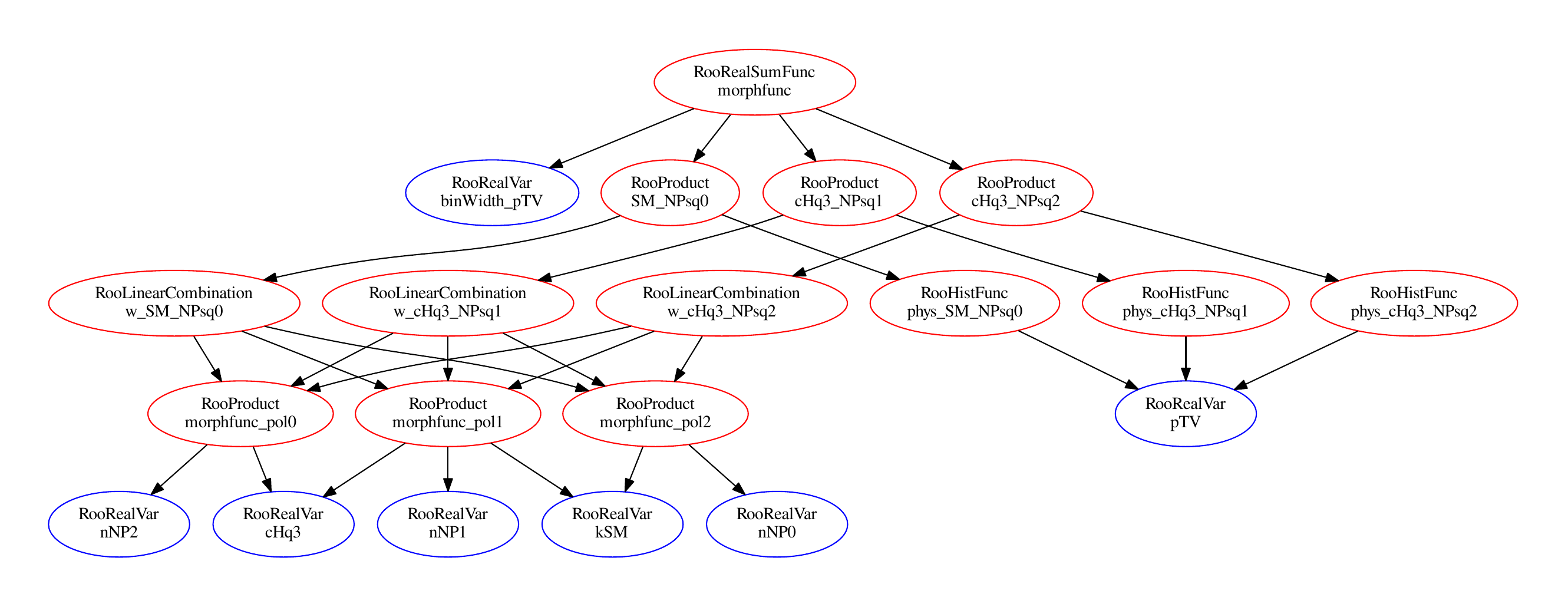}
    \caption{Tree structure of dependency graph of the underlying \rf objects in the morphing function used in the 1D example.}
    \label{fig:morph_example_setup}
\end{figure}

\subsubsection*{Example use case with one morphing parameter}
\label{sec:1dexample}
\begin{figure}[h!]
    \centering
    \includegraphics[width=1.0\textwidth]{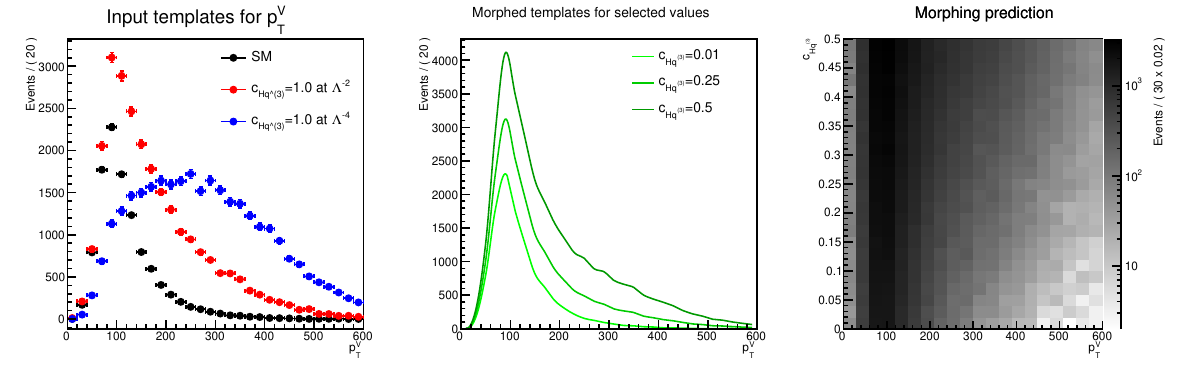}
    \caption{Example of the Lagrangian morphing for a one parameter case. The input distributions are shown on the left. The morphing prediction for select parameter values is shown in the center. The continuous description of the morphing function as a function of the parameter is shown in right.}
    \label{fig:morph_example1}
\end{figure}

A minimal example of the Lagrangian morphing use case is when one
parameter affects one distribution.  In this example, a $2\to2$
process is modelled following the derivation in
Eq.\,\eqref{eq:me-decomposed-2to2} and
\eqref{eq:xsec-decomposed-2to2}. The process in question is $pp\to WH$
with $H\to bb$, generated with \textsc{MadGraph5} \cite{mg5}, and a
single non-\acr{SM} operator ($\Op_{\BSM}$)
$\Op^{(3)}_{Hq}$ is chosen from the Warsaw basis \cite{warsaw-basis}, using
the \texttt{SMEFTsim} model \cite{smeftsim,smeftsim-guide} where the single
insertion of \acr{EFT} operators per diagram is considered. This corresponds to the scenario given by Eq.\,\eqref{eq:me-decomposed} with $(g_{SM},g_{\BSM}) = (1,\texttt{cHq3})$.  To construct the morphing function for this case three non-degenerate input distributions are required. The sample provided 
as inputs to the morphing are the \acr{SM} contribution ($\propto g_{SM}^2$), the interference of
the operator with \acr{SM} ($\propto g_{SM}g_{BSM}$), and the squared order contribution of the operator
for $c^{(3)}_{Hq}=1.0$ ($\propto g_{BSM}^2$) generated using the $\texttt{NP}^{\wedge} \texttt{2==x}$
syntax where x is 0,1, or 2 respectively. The computational graph of \texttt{RooFit} function objects built by the morphing function
is shown in Fig.~\ref{fig:morph_example_setup}. The input
distributions as well as the predictions computed by the morphing
functions are shown in Fig.~\ref{fig:morph_example1}.

The code snippet required to perform this computation using
\texttt{RooLagrangianMorphFunc} is shown in Listing \ref{lst:1d}.  The
corresponding \texttt{RooWorkspace} factory interface usage, which can
be used as an alternative means to the same end, is shown in Listing
\ref{lst:1dfac}.

\begin{lstfloat}[h!]
  \caption{Setup for a simple, one-dimensional morphing. The \text{RooFit} classes shown in the diagram in Fig.~\ref{fig:morph_design} are color-coded here accordingly.}\label{lst:1d}
  \begin{Verbatim}[commandchars=\\\{\},frame=single]
\textcolor{lightgray}{// usage of RooLagrangianMorphFunc for simple case}
\textcolor{lightgray}{// prepare inputs}

\textcolor{lightgray}{// necessary strings for inputs to morphing function}
std::string infilename = "inputs/input_histos.root";
std::string obsname = "pTV";
std::vector<std::string> \textcolor{YellowOrange}{samples} = \{"SM_NPsq0","cHq3_NPsq1","cHq3_NPsq2"\};

\textcolor{lightgray}{// create relevant parameters}
\textcolor{Cyan}{RooRealVar} cHq3("cHq3","cHq3",0,-10,10);
\textcolor{Cyan}{RooRealVar} sm("SM","SM",1);
\textcolor{lightgray}{// Set NewPhysics order of parameter}
cHq3.setAttribute("NewPhysics",true);

\textcolor{lightgray}{// Setup config object}
RooLagrangianMorphFunc::Config config;
config.couplings.add(cHq3);
config.couplings.add(sm);
config.fileName = infilename.c_str();
config.observableName = obsname.c_str();
config.folderNames = samples;

\textcolor{lightgray}{// Setup morphing function}
\textcolor{ForestGreen}{RooLagrangianMorphFunc} morphfunc("morphfunc","morphfunc", config);
\end{Verbatim}

\end{lstfloat}

\begin{lstfloat}[h!]
  \caption{Setup for a simple, one-dimensional morphing using the \texttt{RooWorkspace} factory interface.}\label{lst:1dfac}
  \begin{Verbatim}[commandchars=\\\{\},frame=single]
\textcolor{lightgray}{//RooLagrangianMorphFunc in the RooWorkspace factory interface}
ws.factory("lagrangianmorph::morph(
                   $fileName('inputs/input_histos.root'),
                   $observableName('pTV'),
                   $couplings(\{cHq3[0,1],SM[1]\}),
                   $NewPhysics(cHq3=1),
                   $folders(\{'SM_NPsq0','cHq3_NPsq1','cHq3_NPsq2'\}))");
\end{Verbatim}
\end{lstfloat}  

\subsubsection*{Multiple parameter use case}
\begin{figure}[h!]
    \centering
    \includegraphics[width=1.0\textwidth]{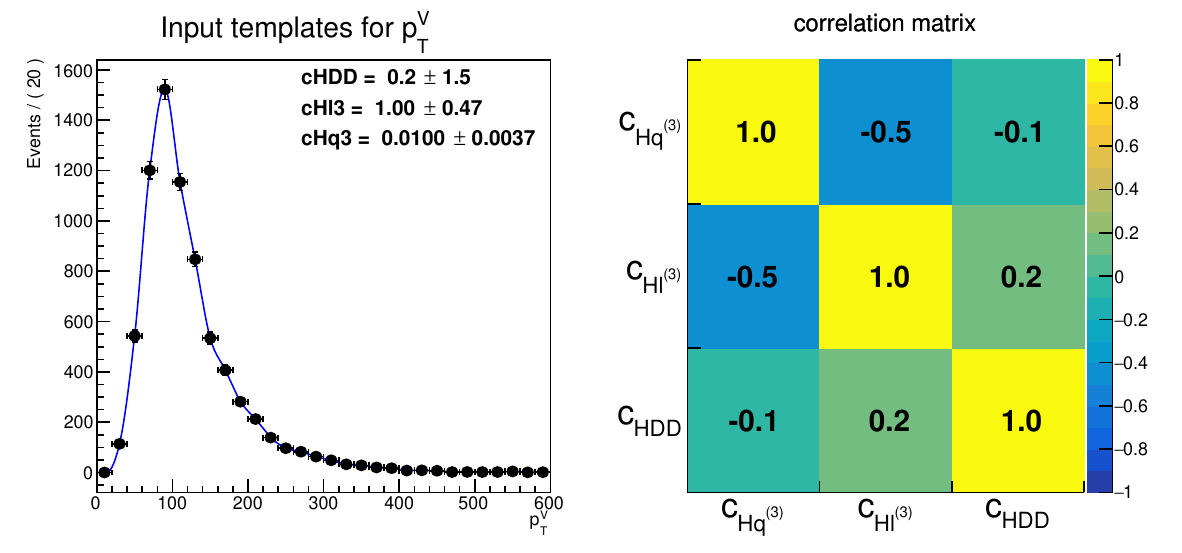}
    \caption{Example of the Lagrangian morphing for a multi-parameter use case. The figure shows the comparison of a fit to a pseudo dataset generated at ${c_{HDD}=0.2,c^{(3)}_{Hl}=1.0,c^{(3)}_{Hq}}=0.01$ as well as the correlation matrix.}
    \label{fig:morph_example2}
\end{figure}

The usage \texttt{RooLagrangianMorphFunc} class can be extended to
handle arbitrary complexity in parameters and observable distributions
simultaneously modelling multiple 1D distributions. This example uses
the same process as Section \ref{sec:1dexample}, but this time, three
different operators are introduced: $\Op^{(3)}_{Hq}$,
$\Op^{(3)}_{Hl}$, and $\Op_{HDD}$ with three corresponding Wilson 
coefficients as parameters $c^{(3)}_{Hq}$, $c^{(3)}_{Hl}$, and $c_{HDD}$.
Following Eq.\,\eqref{eqn:nsamples} 10 input samples are required  
as input templates to the morphing samples for this example.
The 10 samples used here as inputs correspond to four types,
\begin{itemize}
    \item \acr{SM} - the sample of corresponding to the $g_{\acr{SM}}^2$ Eq.\,\eqref{eq:me-decomposed} generated using $\texttt{NP}^{\wedge} \texttt{2==0}$.
    \item \acr{SM}-\acr{BSM} Interference  - three samples of the form $g_{BSM}g_{SM}$ generated using $\texttt{NP}^{\wedge} \texttt{2==1}$ and by setting one of {$c^{(3)}_{Hq}$, $c^{(3)}_{Hl}$, $c_{HDD}$} to $c=1.0$ and the remaining ones to $c=0.0$.
    \item \acr{BSM} Square - three samples of the form $g_{\acr{BSM}}^2$ generated using $\texttt{NP}^{\wedge} \texttt{2==2}$ and by setting one of {$c^{(3)}_{Hq}$, $c^{(3)}_{Hl}$, $c_{HDD}$} to $c=1.0$ and the remaining ones to $c=0.0$.
    \item \acr{BSM}-\acr{BSM} Interference  - three samples including the contribution of $g_{\acr{BSM}_i}g_{BSM_j}, g_{\acr{BSM}_i}^2,  g_{\acr{BSM}_j}^2$ $(i\neq j)$ generated using $\texttt{NP}^{\wedge} \texttt{2==2}$ and by setting two of {$c^{(3)}_{Hq}$, $c^{(3)}_{Hl}$, $c_{HDD}$} to $c=1.0$ and the remaining ones to $c=0.0$.
\end{itemize}
The morphing distribution for a given parameter point is shown in Fig.~\ref{fig:morph_example2}.

\subsection*{Other parametrization scenarios}

\subsubsection*{Ratio of Polynomials}
As discussed in Section \ref{sec:linear} the parameterisation for the
cross-section can be modelled as a rational polynomial in the
parameters when modelling separately the \acr{EFT} behaviour of the
of a narrow s-channel resonance and branching ratio as discussed in \ref{eq:ratiopoly}. 
The total cross-section in this scenario can be represented with the 
new \texttt{RooRatio} class which provides the
option to build products and ratio of individual \texttt{RooAbsReal}
function objects. To this end, the \texttt{RooLagrangianMorphFunc}
implementation provides a dedicated \texttt{makeRatio} method to
construct the ratio of morphing functions, which is showcased in 
Listing \ref{lst:ratio}. The same can
also be achieved with existing morphing functions. The syntax
for this is shown in Listing \ref{lst:ratiofac}, using the
\texttt{RooWorkspace} factory interface.

\begin{lstfloat}[h!]
  \caption{Usage of the \texttt{makeRatio} method to construct a ratio of morphing functions.}\label{lst:ratio}    
  \begin{Verbatim}[commandchars=\\\{\},frame=single]
\textcolor{lightgray}{// RooRatio to setup up ratio of morphing functions}
\textcolor{lightgray}{// morphfunc_prod --> function to model production EFT}
\textcolor{lightgray}{// morphfunc_partial_width --> function to model partial width EFT}
\textcolor{lightgray}{// morphfunc_total_width --> function to model total width EFT}

\textcolor{lightgray}{// Setup numerator and denominator functions}
RooArgList nr(morphfunc_prod,morphfunc_partial_width);
RooArgList dr(morphfunc_total_width);

\textcolor{lightgray}{// create ratio}
auto ratio = RooLagrangianMorphFunc::makeRatio("ratio","ratio", nr, dr);
\end{Verbatim}

\end{lstfloat}

\begin{lstfloat}[h!]
  \caption{Creation of a ratio of two morphing functions with the \texttt{RooWorkspace} factory interface.}\label{lst:ratiofac}      
  \begin{Verbatim}[commandchars=\\\{\},frame=single]
\textcolor{lightgray}{// RooRatio to setup up ratio of morphing functions}
\textcolor{lightgray}{// in RooWorkspace factory interface}
ws.factory("Ratio::ratio(\{morphfunc_prod,morphfunc_partial_width\},
                           \{morphfunc_total_width\});
\end{Verbatim}

\end{lstfloat}  

\subsubsection*{Taylor Expansion}

The \texttt{RooPolyFunc::taylorExpand} method provides the Taylor
expansion of any function with respect to a set of parameters with subsequent 
truncation at either first or second order for improved interpretability of the result.
Listings \ref{lst:taylor} and \ref{lst:taylorfac} 
shows the usage of the truncated Taylor expansion formalism in \texttt{C++} and 
the workspace factory language respectively.

\begin{lstfloat}[h!]
  \caption{Usage of the automatic Taylor expansion to obtain a morphing function at quadratic order based on a ratio of cross sections and branching ratios.}\label{lst:taylor}  
  \begin{Verbatim}[commandchars=\\\{\},frame=single]
\textcolor{lightgray}{// RooRatio to setup up ratio of morphing functions}
\textcolor{lightgray}{// in RooWorkspace factory interface}
\textcolor{lightgray}{// prodxBR --> RooRatio to model EFT in production & branching ratio}
\textcolor{lightgray}{// truncating second-order Taylor expansion of prodxBR w.r.t (par1, par2)}
\textcolor{lightgray}{// around 0.0 for both parameters}
auto prodxBR_taylor = RooPolyFunc::taylorExpand("prodxBR_taylor",
                                                "prodxBR_taylor",
                        \textcolor{lightgray}{// function to Taylor expand}
                                                prodxBR, 
                        \textcolor{lightgray}{// parameters to Taylor expand around}
                                               RooArgList(par1,par2), 
                        \textcolor{lightgray}{// expand around parameter coordinate,}
                        \textcolor{lightgray}{// one value of all parameters (0.0)}
                        \textcolor{lightgray}{// or vector (\{0.0,0.0\})} 
                                                0.0, 
                        \textcolor{lightgray}{//truncation order of expansion}
                                             2);
\end{Verbatim}

\end{lstfloat}

\begin{lstfloat}[h!]
  \caption{Usage of the automatic Taylor expansion with the \texttt{RooWorkspace} factory interface.}\label{lst:taylorfac}  
  \begin{Verbatim}[commandchars=\\\{\},frame=single]
\textcolor{lightgray}{// RooRatio to setup up ratio of morphing functions}
\textcolor{lightgray}{// in RooWorkspace factory interface}
ws.factory("taylorexpand::prodxBR_taylor(prodxBR,\{par1,par2\},0.0,2);
\end{Verbatim}
\end{lstfloat}

\section*{Acknowledgments}
We gracefully acknowledge funding from the Union European Union’s Horizon 2020 research and innovation programme, call \acr{H2020-MSCA-ITN-2017}, under Grant Agreement n.~765710.

We thank the \texttt{ROOT} development team at \acr{CERN} for its support in integrating this software their environment.

\bibliographystyle{custombib}
\bibliography{main}

\end{document}